\documentclass[aps,secnumarabic,showpacs,nobibnotes,showkeys,superscriptaddress,
amssymb,prd,epsfigs,amsfonts,nofootinbib,amsmath,apsfonts]{revtex4}
\def\prl{{\em Phys. Rev. Lett. }}
\def\prc{{\em Phys. Rev. {\bf C} }}

\def\nima{{\em Nucl. Instr. and Meth. Phys. {\bf A} }}
\def\npa{{\em Nucl. Phys. {\bf A}}}
\def\npb{{\em Nucl. Phys. {\bf B}}}
\def\epjc{{\em Eur. Phys. J. {\bf C}}}
\def\plb{{\em Phys. Lett. {\bf B}}}
\def\mpla{{\em Mod. Phys. Lett. {\bf A}}}
\def\pr{{\em Phys. Rep.}}
\def\zpc{{\em Z. Phys. {\bf C}}}

\usepackage{graphicx}
\usepackage{dcolumn}
\usepackage{bm}

\begin{document}
\title{\Large\bf Transverse Energy Measurement in $\sqrt{s_{NN}} = 62.4 $ GeV Au+Au
       Collisions at RHIC}
\affiliation{Institute of Physics, Bhubaneswar, India}
\affiliation{University of Sao Paulo, Brazil}
\affiliation{Variable Energy Cyclotron Centre, Kolkata, India}

\author{Raghunath Sahoo$^*$}\affiliation{Institute of Physics, Bhubaneswar, India}
\author{Subhasis Chattopadhaya}\affiliation{Variable Energy Cyclotron Centre, 
Kolkata, India}
\author{Alexandre A. P. Suaide}\affiliation{University of Sao Paulo, Brazil}
\author{Marcia Maria de Moura}\affiliation{University of Sao Paulo, Brazil} 
\author{D.P. Mahapatra}\affiliation{Institute of Physics, Bhubaneswar, India}

\collaboration{for the STAR Collaboration}\noaffiliation


\footnote{$^*$  Corresponding author: Raghunath Sahoo$^1$ (email: raghu@iopb.res.in)}

\begin{abstract}
  The transverse energy distributions ($E_{T}$)
have been measured for Au + Au collisions at
$\sqrt{s_{NN}} = 62.4$ GeV by the STAR experiment
at RHIC. They have been obtained from two measurements,
the hadronic transverse energy ($E_{T}^{had}$) and
the electromagnetic transverse energy($E_{T}^{em}$).
$E_{T}^{had}$ has been measured from the tracks
obtained by Time Projection Chamber (TPC) excluding
the electrons and positrons. $E_{T}^{em}$ has been obtained
by the STAR Barrel Electromagetic Calorimeter (BEMC)
which measures the energy of electrons, positrons and
photons. The measure of transverse energy  gives an
estimate of the energy density of the fireball produced
in heavy ion collisions. $E_{T}$ per participant pair
gives information about the  production  mechanism of
particles.  

\end{abstract}
\maketitle
\section{Introduction}

The quest for understanding of the possible formation 
and existence of the quark-gluon plasma (QGP), the 
deconfined phase of quarks and gluons, has been a major 
area of heavy-ion research during the last couple of 
decades. The study of high energy nuclear collisions 
at the Relativistic Heavy Ion Collider (RHIC)[1] has 
opened a new domain in the exploration of strongly interacting matter at 
very high enegy density. High temperature and densities 
may be generated in the most central nuclear collisions
at RHIC, creating conditions in which a phase of 
deconfined quarks and gluons may exist [2,3]. The 
understanding of the early phase of the fireball 
produced in nuclear collisions requires the study of 
observables like the energy and momentum, produced 
transversely to the beam direction viz $E_T$ and $p_T$.
The number of charge particles produced and $E_T$ are 
closely related to the collision geometry and are of importance in 
understanding the global properties of the system formed
during the collision. Scattering of the partonic 
constituents of the incoming nuclei in the initial 
phase added to the rescattering of the produced 
partons and hadrons results in generation of $E_T$[4,5]. 
If the fireball of produced quanta breaks apart quickly 
without significant reinteraction, the observed 
$dE_T/dy$ will be the same as that produced by the 
initial scattering. However, with interaction among
the produced quanta, the system can achieve equilibrum
at a very early state, after which it can expand resulting
in a lowering of $dE_T/dy$[6,7]. To some extent this will
be compensated by any transverse hydrodynamic flow[8].
Gluon saturation can delay the onset of the above 
hydrodynamic flow reducing the effective pressure and 
thereby reducing the difference in the initially produced
and the observed $E_T$[9].\\
Here, we present preliminary results on $E_T$, produced 
in 62.4 GeV Au+Au collisions at RHIC. We present both the 
hadronic and the electromagnetic components of $E_T$, which 
were measured independently by the STAR detector using the 
Time projection Chamber (TPC) and the Barrel Electromagnetic 
Calorimeter (BEMC). A brief description of the STAR experiment 
with the detectors used together with the data analysis methods
employed is given in Section-2. In Section-3 we present
the results (STAR Preliminary) with conclusion presented
in Section-4.
   
\section{STAR Experiment and Data Analysis}

STAR[10], is an azimuthally symmetric, large acceptance 
solenoidal detector comprising of several detector
subsystems. The subsystems relevant for this analysis 
are a large TPC located inside a 0.5 T solenoidal magnet, 
the BEMC and two zero-degree calorimeters (ZDCs) for 
event selection. The BEMC[11] is a lead-scintillator 
sampling electromagnetic calorimeter with equal volumes 
of lead and scintillator. The full Barrel corresponds to 
120 modules. Each module is composed of 40 towers (20 towers 
in $\eta$ by 2 towers in $\phi$), constructed to project 
to the center of the STAR detector. For the 2004 run, 
only half of the Barrel was instrumented, corresponding 
to a pseudorapidity coverage of $0 < \eta <1$ with full 
azimuthal symmetry. The transverse dimensions of a tower are 
approximately $10 \times 10 ~cm^2 $, which at the radius 
of the front face of the detector coresponds to a phase 
space interval $ (\Delta\eta, \Delta\phi)$ of (0.05, 0.05). 
Each tower has a depth of 21 radiation lengths($X_0$) 
corresponding to one interaction length for a hadron. 
The BEMC with a radius of 2.3 m sits inside the STAR 
solenoidal magnet. The electromagnetic energy resolution 
of the BEMC is $\delta E/E \sim 16 \%  / \sqrt{E(GeV)}$. \\
The TPC[12] is the primary STAR detector used for the 
event reconstruction. It is a gas chamber, 4.2 m long 
with inner and outer radii of 50 and 200 cm respectively, 
placed in an uniform magnetic field of 0.5 T. The particles 
passing through the active gas medium release secondary 
electrons that drift to the readout end caps at both ends 
of the chamber. The readout system is based on multiwire 
propotional counters, with readout pads. There are 45 pad 
rows between the inner and outer radii of the TPC. The 
induced charge from the electrons is shared over several 
adjacent pads. The TPC provides up to 45 independent 
spatial and specific ionization $ dE/dx $ measurements. 
The $ dE/dx $ measurements, along with the momentum 
measurement from the bending of the tracks inside the magnetic 
field, determine the particle mass within limited kinematic 
regions. The TPC covers a pseudorapidity region with 
$|\eta| < 1.8 $ with full azimuthal coverage.\\
The event trigger consisted of the coincidence of signals 
from two ZDCs, located at $ \theta < 2 $ mrad about the beam 
down stream of the first accelerator dipole magnet and are sensitive 
to spectactor neutrons. These calorimeters provide a minimum 
bias trigger which, after collision vertex reconstruction, 
coresponds to $ 97 \pm3\% $ of the geometric cross section 
$ \sigma ^{Au+Au}_{geom} $. The events are analyzed in centrality 
bins based on the TPC charged particle multiplicity in 
$ |\eta| < 0.5 $. \\
The present analysis is based on the minimum bias Au+Au  
collisions data  at $\sqrt{s_{NN}} = 62.4$ GeV, taken by 
STAR in the 2004 RHIC run. Here the TPC acceptance is limited 
by the BEMC. The TPC track quality cuts include a) z-coordinate 
(longitudinal axis) selection of collision vertex within 30 cm 
of the TPC center and b) a minimum TPC track space point cut of 10.

\subsection{Hadronic Transverse Energy, $E_{T}^{had}$}

The hadronic part of the transverse energy as measured from 
the momentum analysed TPC tracks is defined as\\
\begin{eqnarray}
E_{T}^{had} = \sum_{tracks} E_{track}sin\theta,
\end{eqnarray}
where the sum runs for all hadrons produced in the collision, 
except $\pi^0$, $\eta$ and other long-lived neutral hadrons. 
$\theta$ is the polar angle with 
respect to the beam axis and the collision vertex position. 
The Hadronic energy, $E_{track}$, is defined as [13,14]

\[E_{track} = \left\{ \begin{array}{lll}
	\sqrt{p^2 + m^2}-m, & \mbox{for nucleons}\\
	\sqrt{p^2 + m^2}+m, & \mbox{for anti-nucleons}\\
	\sqrt{p^2 + m^2}, & \mbox{for all other particles}
	\end{array}
	\right. \]

With the above definition,
\begin{eqnarray}
E_{T}^{had} = C_0 \sum_{tracks} C_1(ID,p)E_{track}(ID,p)sin\theta
\end{eqnarray}
where the sum includes all the primary tracks in BEMC acceptance. 
Here, $C_0$ is a correction factor defined as
\begin{eqnarray}
C_0 = \frac{1}{f_{acc}}\frac{1}{f_{pT^{cut}}}\frac{1}{f_{neutral}}
\end{eqnarray}
where $f_{acc}$ is the acceptance correction, $f_{neutral}$ is the 
correction for long-lived neutral hadrons not measured by the TPC,
$f_{pT^{cut}}$ corresponding to the TPC low momentum cutoff. 
The factor $C_1(ID,p)$ is defined as
\begin{eqnarray}
C_1(ID,p) = f_{bg}(pT)\frac{1}{f_{notID}}\frac{1}{eff(pT)}
\end{eqnarray}
which includes the uncertainty in the particle ID determination, 
$f_{notID}$, the momentum dependent tracking efficiency, eff(pT) 
and the momentum dependent background, $f_{bg}(pT)$. Details of 
the procedure for finding out the correction factors are given
elsewhere[13]. 

\subsection{Electromagnetic Transverse Energy, $E^{em}_T$}

$E^{em}_T$ is measured from the BEMC tower hits corrected for 
the hadronic contaminations in the calorimeter. It is defined as
\begin{eqnarray}
E_T^{em} = \sum_{towers}E_{tower}^{em}sin(\theta_{tower})
\end{eqnarray}
where $E_{tower}^{em}$ is the electromagnetic energy measure 
in an BEMC tower and $\theta_{tower}$ is the polar angle of 
the center of the tower relative to the beam axis and the 
collision vertex position. Experimentally, $E_T^{em}$ is given 
by
\begin{eqnarray}
E_T^{em} = \frac{1}{f_{acc}}\sum_{towers}(E_{tower}-
\Delta E_{tower}^{had})sin(\theta_{tower})
\end{eqnarray}
where, $f_{acc}$ is the acceptance correction, $E_{tower}$ is 
the energy measured by an BEMC tower and $\Delta E_{tower}^{had}$ 
is the total correction for each tower to exclude the hadronic 
contributions. $\Delta E_{tower}^{had}$ is given by
\begin{eqnarray}
\Delta E_{tower}^{had} = \frac{1}{f_{neutral}}\sum_{tracks}
\frac{f_{elec}(pT)}{eff(pT)}\Delta E(p,\eta,d),
\end{eqnarray}
where, $\Delta E(p,\eta,d)$ is the energy deposited by a 
track projected on a BEMC tower as a function of it's momentum 
$p$, pseudorapidity $\eta$ and distance $d$ from the center of 
the tower to the track hit point. $f_{elec}(pT)$ is the 
correction to exclude electrons that are misidentified as 
hadrons. eff(pT) is the track reconstruction efficiency and 
$f_{neutral}$ is the contribution to exclude the long-lived 
neutral hadron contribution.

\section{Results}

Assuming the correction factors used in this analysis, 
to be the same as those obtained for 200 GeV Au+Au 
collisions[13], we have determined the minimum bias 
distributions of the total tranverse energy, $E_T$ and it's 
components ($E_T^{had}$ and $E_T^{em}$) separately. These
are shown in Fig.1.(a)-(c)

\begin{figure}[htbp]
\includegraphics[width=2.3in]{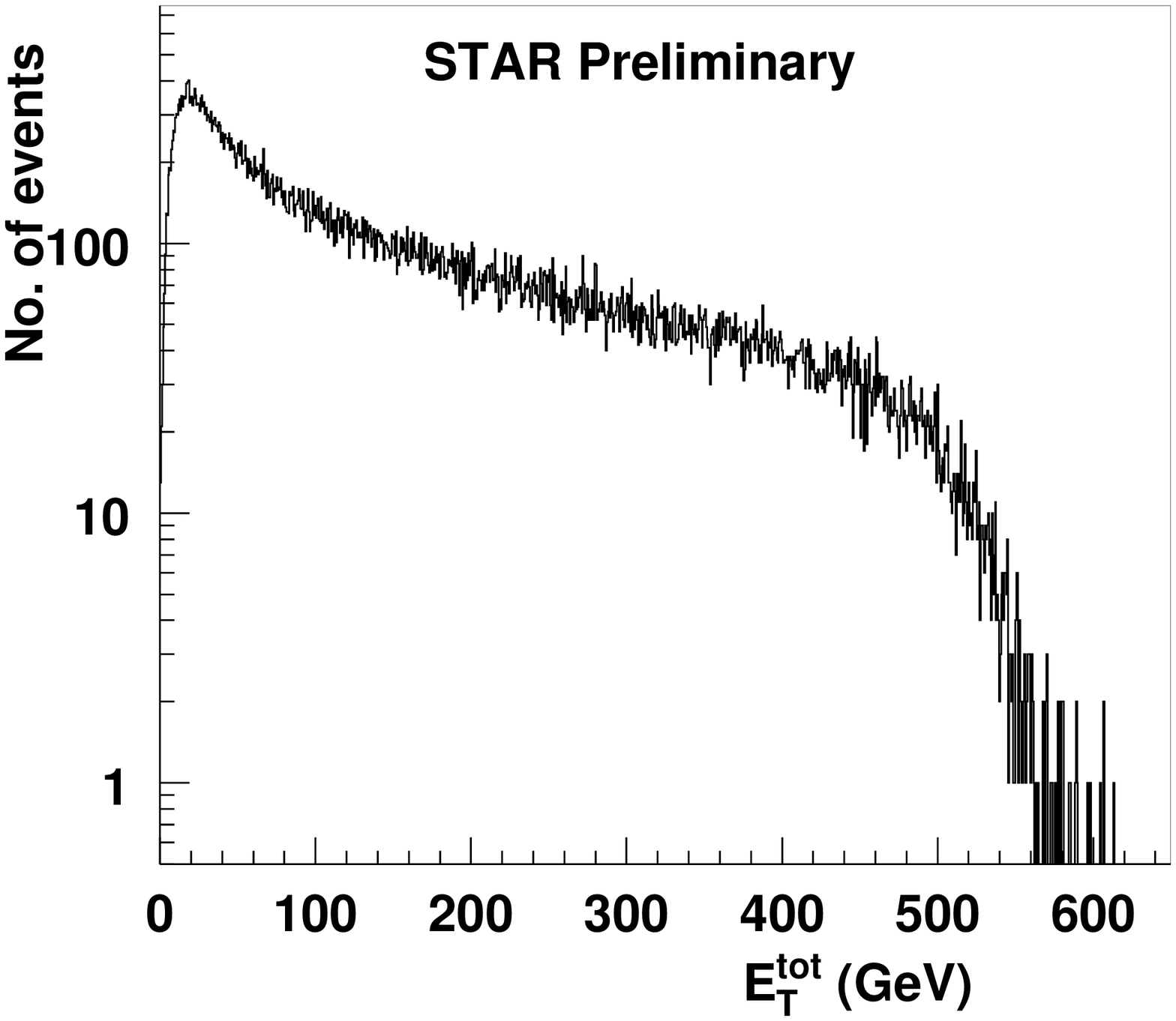}
\includegraphics[width=2.3in]{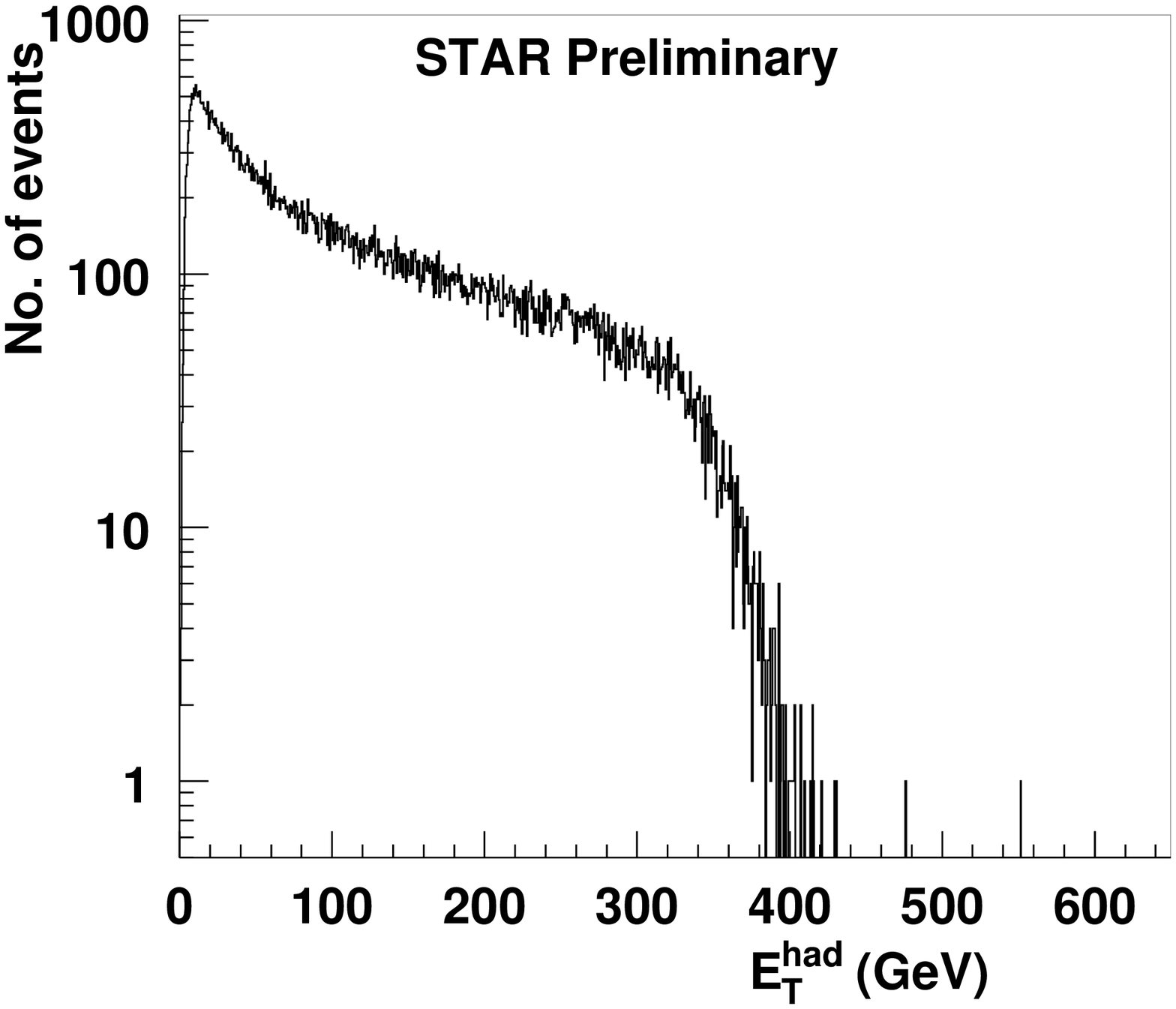}
\includegraphics[width=2.3in]{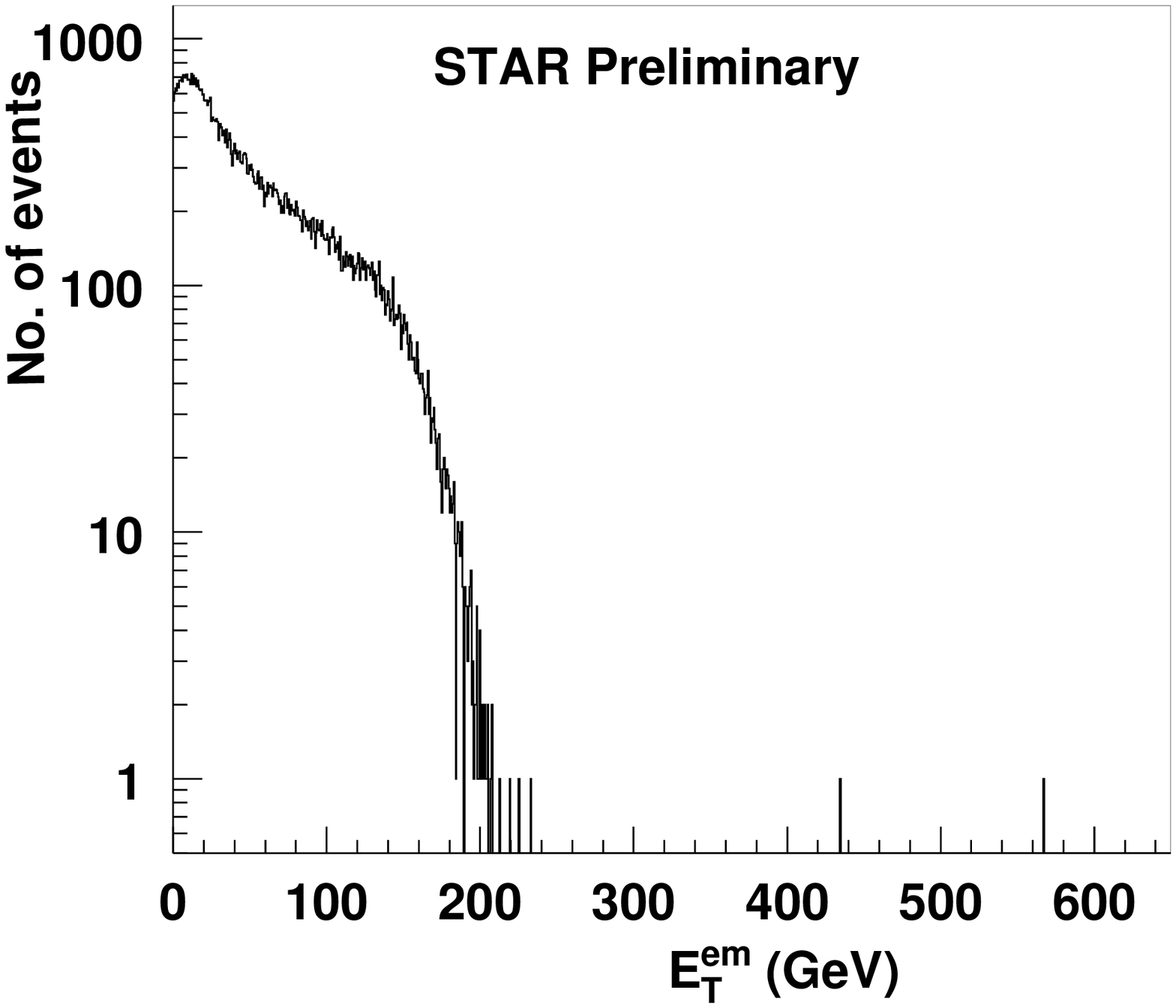}
\caption{Minimum bias distributions of (a) total transverse energy, 
(b) hadronic transverse energy and (c) electromagnetic transverse energy, 
for $\sqrt{s_{NN}} = 62.4$ GeV Au+Au collisions.}
\end{figure}

In Fig. 2(a), we present $dE_T/dy$ per participant pair for 
the top 5\% central collisions, together with results from 
other experiments from AGS to RHIC[15-18]. The $dE_T/dy$ values
for this analysis, were calculated from $dE_T/d\eta$ using a factor of 1.18 
obtained from HIJING simulation to convert $\eta$ to $y$ phase 
space. We obtained a value of 3.05 $\pm$ ~0.05~(stat) GeV, consistent 
with an overall logarithimic growth of $dE_T/dy/(0.5N_{part})$ 
with $\sqrt{s_{NN}}$. For the same top 5\% central events, 
$<E_T>_{5\%}$ as obtained from $<dE_T/d\eta|_{\eta=0.5}>$ for 
full azimuthal coverage and one unit $\eta$ interval, is found
to be 450 $\pm $6~(stat) GeV. Further, we have also determined the 
spatial energy density produced in the collision using the 
Bjorken formula[19]
\begin{eqnarray}
\epsilon_{Bj} = \frac{dE_T}{dy}\frac{1}{\tau_0 \pi R^2}
\end{eqnarray}
where, $dE_T/dy$ is the primordial rapidity density of 
transverse energy, R is the transverse system size given by 
$ R = R_0 A^{\frac{1}{3}}$ and $\tau_0$ is the formation time. 
Assuming $\tau_0$ = 1 fm/c, we get $\epsilon_{Bj} = 3.46 \pm 0.05~ GeV/fm^3$.
This energy density is significantly higher than the energy 
density $\sim$ $1GeV/fm^3$ as required for the transition to a 
deconfined quark gluon plasma, predicted by lattice QCD [20]. 
This estimate is based on the assumption that local equilibrium
has been achieved at $\tau \sim 1 fm/c$ and then the system 
expands hydrodynamically.\\

\begin{figure}[htbp]
\includegraphics[width=3.5in]{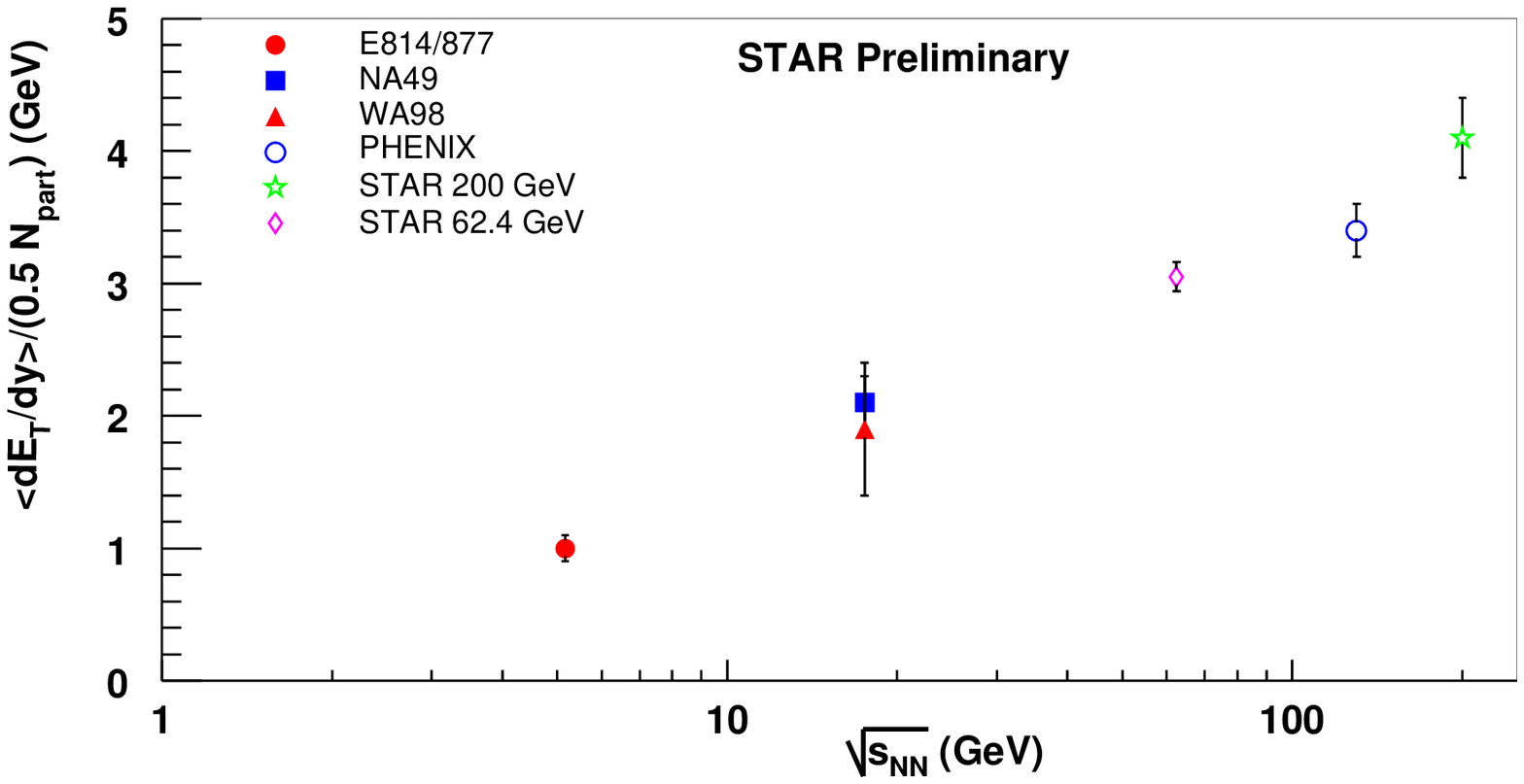}
\includegraphics[width=3.5in]{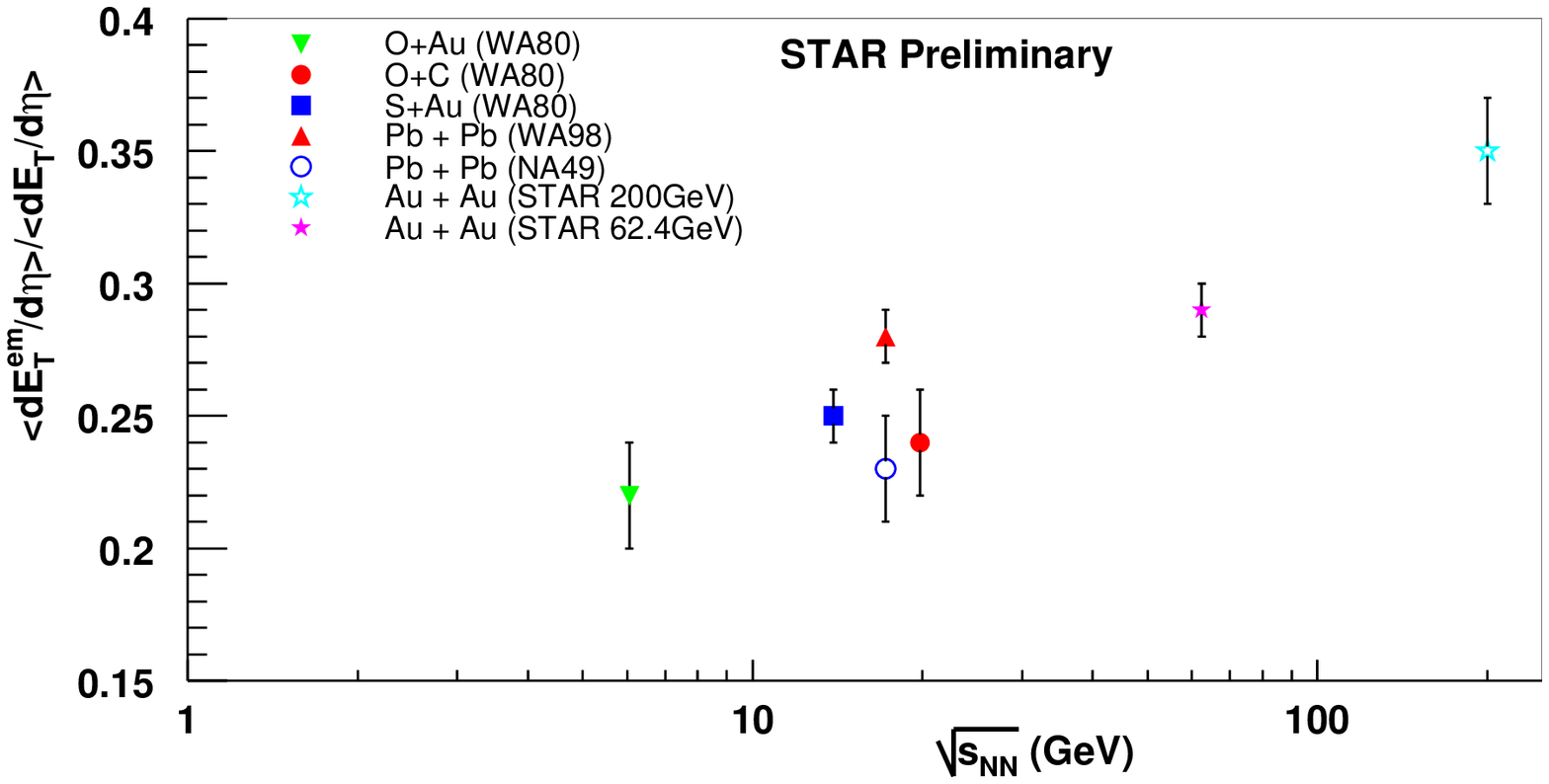}
\caption{(a) $(dE_T/dy)/(0.5 N_{part}) ~Vs \sqrt{s_{NN}}$ for 
central events, (b) Collision energy dependence of electromagnetic 
fraction of total $E_T$ as given by $<dE_T^{em}/d\eta>/<dE_T/d\eta>$ 
for a number of systems from SPS to RHIC for central events}
\end{figure}

In Fig. 2(b), we have shown the electromagnetic fraction of 
the total transverse energy for the top 5\% central events, 
as a function of the center of mass energy from SPS[21,22] 
to RHIC. This is seen to increase slowly when we go from AGS 
to RHIC. At 62.4 GeV this value is $0.29 \pm 0.01$~ (stat). As discussed 
in ref.[13], the observed electromagnetic fraction of the 
total transverse energy is strongly dependent on the baryon to 
meson ratio. At very high energy it is expected that virtually 
all the $E_T$ will be carried by mesons, as an almost baryon 
free region is expected to be created in the central rapidity
region, while at lower SPS energies, baryon dominance results 
in a much smaller electromagnetic fraction. A very high value 
of the electromagnetic fraction of total transverse energy is 
expected in case of a long-lived deconfined phase, due to an 
excess yield of photons[23]. Based on the present ratio
of $0.29 \pm 0.01$~ (stat) it is very difficult to conclude 
anything regarding the formation of a deconfined QGP phase.

\section{Conclusions}

In the present work we have reported priliminary STAR results
on $E_T$ within $0 < \eta <1$ for Au+Au collissions at 
$\sqrt{s_{NN}}= 62.4 $ GeV. For top 5\% central events
$<E_T>$ has been estimated to be 450 $\pm 6$~(stat) GeV. 
The value of $<dE_T/dy>/(0.5 N_{part})$ has been found to be
$3.05 \pm 0.05$~(stat) GeV. Knowing that the observed $E_T$ is lower than the 
initial values [6,9,24], the present value of $3.05 \pm 0.05$~(stat) GeV 
may be considered as a lower bound only. The initial energy 
density estimated within the framework of boost-invariant 
hydrodynamics, as given by $\epsilon_{Bj}$, has been found to 
be $3.46 \pm 0.05 ~(stat) GeV/fm^3 $ which is well above that required for the 
deconfinement phase transition as predicted by lattice QCD[20]. 
The electromagnetic fraction of the total transverse energy 
for central events is found to be $0.29\pm0.01$~(stat), consistent 
with the fact that the final state is dominated by mesons. Finally,
this method of independent measurement of $E_T^{em}$ and $E_T^{had}$
gives an unique opportunity to study event by event fluctuations in 
these observables and in their ratios. The correction factors and the
systematic errors are yet to be calculated for this analysis. 

\begin{acknowledgments}

We thank the RHIC Operations Group and RCF at BNL, and the
NERSC Center at LBNL for their support. This work was supported
in part by the HENP Divisions of the Office of Science of the U.S.
DOE; the U.S. NSF; the BMBF of Germany; IN2P3, RA, RPL, and
EMN of France; EPSRC of the United Kingdom; FAPESP of Brazil;
the Russian Ministry of Science and Technology; the Ministry of
Education and the NNSFC of China; IRP and GA of the Czech Republic,
FOM of the Netherlands, DAE, DST, and CSIR of the Government
of India; Swiss NSF; the Polish State Committee for Scientific 
Research; STAA of Slovakia, and the Korea Sci. \& Eng. Foundation.
\end{acknowledgments}

\end{document}